\documentclass[prl,preprint,amsmath,longbibliography]{revtex4-1}
\usepackage{graphicx}
\usepackage{amsfonts}
\usepackage{bbm}
\usepackage{color}
\usepackage[colorlinks]{hyperref}
\hypersetup{urlcolor=blue}
\def\Z{\ensuremath{\mathcal{Z}}}

\def\P{\ensuremath{\mathcal{P}}}
\def\W{\ensuremath{\mathcal{W}}}
\def\M{\ensuremath{\mathcal{M}}}

\def\id{\ensuremath{\mathbbm{1}}}
\DeclareMathOperator{\Tr}{Tr}
\begin{document}
\title{Quantum state tomography from sequential measurement\\of two variables in a single setup}
\author{Antonio \surname{Di Lorenzo}}
\affiliation{Instituto de F\'{\i}sica, Universidade Federal de Uberl\^{a}ndia, 38400-902 Uberl\^{a}ndia, Minas Gerais, Brazil}
\affiliation{CNR-IMM-UOS Catania (Universit\`a), Consiglio Nazionale delle Ricerche,
Via Santa Sofia 64, 95123 Catania, Italy}
\begin{abstract}
We demonstrate that the task of determining an unknown quantum state 
can be accomplished efficiently by making a sequential measurement  
of two observables $\Hat{A}$ and $\Hat{B}$, 
the eigenstates of which form bases connected by a discrete Fourier transform. 
The state can be pure or mixed, the dimension of the Hilbert space and the coupling strength are arbitrary, 
and the experimental setup is fixed.
The concept of Moyal quasicharacteristic function is introduced for finite-dimensional Hilbert spaces. 
\end{abstract}
\maketitle
\begin{widetext}
\break
\end{widetext}
\section{Introduction}
A colleague has challenged you: she has built a black box from which, upon the pressing of a button, a quantum system is released. What is the state of the system?
You are not allowed to open the box, nor to measure any of its properties.
You can only measure the quantum system, and repeat as many times as you want. 
This is the essence of quantum state tomography.

The preparation of a quantum system is characterized by a quantum state, which is given by 
the density operator, a positive-definite operator of trace one in a Hilbert space. 
Often, some information about the system is missing, but it could be recovered, in principle, from the environment and from 
the preparing apparatus. When all this information is retrieved, which can be done without disturbing the system in any way, 
the quantum system is described by a pure state, i.e. a density operator of rank one, which can be written as 
$\rho_\mathrm{sys}=|\psi\rangle\langle\psi|$ in terms of a vector $|\psi\rangle$ of the Hilbert space. 
However, in general, this information is lost for all practical purposes, and the system is to be described by 
a density operator of higher rank.
A fundamental question is then, how do we determine the unknown state $\rho_\mathrm{sys}$ of a quantum system? 

Reconstructing the unknown quantum state $\rho_\mathrm{sys}$ is believed to be a difficult task, requiring the separate measurement of several observables. The usual approach is to take the system in the unknown state and 
measure the statistics of an observable $\Hat{A}_1$, then, with a distinct ensemble of identically prepared systems, 
measure another observable $\Hat{A}_2$, etc. The observables $\Hat{A}_1,\Hat{A}_2,\dots,\Hat{A}_n$ needed to reconstruct the quantum state are known as the \emph{quorum}, and they usually number as $d^2$, 
with $d$ the dimension of the Hilbert space, even though some improvement over this number can be achieved \cite{Gross2010}. 
Usually, from each measurement, only the average value is extracted. For instance, to reconstruct the 
state of a spin 1/2 system, the average values $n_j=\langle\sigma_j\rangle$, $j=x,y,z$, are calculated, and the state 
$\rho_\mathrm{sys}=(1+\mathbf{n}\cdot\boldsymbol{\sigma})/2$ is reconstructed. The noise introduced by the detectors 
is then a hindrance. 
However, it is important to notice that the full probability distribution of the output is a function 
(typically, a convolution) of both the initial state of the detector and of $\rho_\mathrm{sys}$. Thus, extracting only one 
number, the average, out of the many repetitions of a measurement is extremely limitative and a waste of useful information. 

Furthermore, the most commonly used statistical tool for the reconstruction of the state is the maximum likelihood 
estimation, which does not take into account the positive-definiteness of the density operator and 
may give rise to rank-deficient estimates. \emph{Ad hoc} corrections are 
often devised to overcome this difficulty. The recently introduced Bayesian \cite{Blume2010} approach has solved 
this last issue, but its adoption is being slow.
We remark that in the Bayesian approach, the maximum likelihood estimate is justified when uniform priors are assumed 
and a particular cost function is postulated \cite{Helstrom1976}. 
In any case, the number of different setups needed for quantum state tomography 
increases with the dimension of the Hilbert space, making the process time-consuming.   

Recently, many schemes based on weak measurement 
\cite{Hofmann2010,Lundeen2012,Fischbach2012,Wu2013}
have been proposed for quantum state tomography. 
Experimental realizations were also demonstrated \cite{Lundeen2011,Salvail2013}. 
However, a distinct disadvantage of such schemes is that on one hand the formulas for the weak measurement are 
approximated, introducing a further uncertainty in the reconstruction, and on the other hand the weak measurement 
relies on postselection, which requires that only a fraction of the data is retained, yielding a reduced efficiency. 

Haapasalo \emph{et al.} \cite{Haapasalo2011} have also pointed out the superiority of phase space methods over the 
weak measurement methods in order to reconstruct the wave-function. This suggests to look for an extension 
of phase space methods to finite dimensional Hilbert spaces. In doing so, we shall propose a generalization of 
the Moyal function \cite{Moyal1949}. 
The justification for this choice is that the Moyal function has revealed itself to be an extremely useful tool for describing 
the statistics of joint and sequential measurements of momentum and position \cite{DiLorenzo2011,DiLorenzo2013a}. 

A promising avenue for efficient quantum state tomography was opened by considering 
measurements in mutually unbiased bases \cite{Ivonovic1981,Johansen2008,Kalev2012}. 
All the proposals of which we are aware, however, require many different setups, at least as many as the dimension of the Hilbert space.

Here, instead, we propose a quantum state tomography scheme consisting in a \emph{single} 
sequential measurement of \emph{arbitrary strength} and relying on an \emph{exact relation} between the initial state of 
the system and the final output of the measurement. The whole statistics of the measurement is used, and the 
unsharpness of the detector is turned into a resource, rather than an obstacle.  
Our scheme uses a particular pair of mutually unbiased bases, the Fourier conjugated bases.  
We demonstrate that there are infinitely many pairs of observables $\Hat{A},\Hat{B}$ that allow the 
reconstruction of an unknown quantum state $\rho_\mathrm{sys}$, be this pure or mixed. Furthermore, by suitably choosing 
the first measured observable $\Hat{A}$, it is possible to obtain the representation of the state, 
$\langle m|\rho_\mathrm{sys}|m'\rangle$, 
in any basis of choice. We recover the results of Ref.~\cite{DiLorenzo2013a} in the limit $d\to\infty$. 
Furthermore, a sequential measurement of position and momentum may lead to a violation of the Heisenberg noise-disturbance 
principle \cite{DiLorenzo2013c} if the detectors are initially in a correlated state. The result provided here applies whether 
the detectors are initially correlated or not. 

A related proposal was made by Leonhardt \cite{Leonhardt1995b,Leonhardt1996}, who introduced a different quantum characteristic function for discrete 
systems (see the appendix for a discussion), and proposed to use Ramsey techniques to transform the quadrature 
observables into energy eigenstates. 
Furthermore, recently Carmeli \emph{et al.} \cite{Carmeli2012} 
have demonstrated that sequential measurements of conjugated observables 
are informationally complete, i.e., for any two different density matrices of the system $\rho_1\neq\rho_2$, the probabilities 
differ, $P(A,B|\rho_1)\neq P(A,B|\rho_2)$. Thus, in principle, there is a one-to-one correspondence between the density matrices 
and the probabilities $P(A,B|\rho)$. The present manuscript provides this correspondence. 
%

%
\section{Preliminary definitions}
We resume the conventions used throughout this paper: 
\begin{itemize}
\item
$d$ integer, dimension of the Hilbert space; 
\item
$S=(d-1)/2$ integer or half-odd ``spin''; 
\item
$m,m'$ integer or half-odd numbers spaced by 1 in the range $[-S,S]$;
\item
$\mu=m-m'$ integer in the range $[1-d,d-1]$;
\item
$\bar{M}=\frac{m+m'}{2}$ integer or half-odd in the range $[-S+|\mu|/2,S-|\mu|/2]$ for fixed $\mu$;
\item $\mathcal{I}$ integers or half-odd numbers in the range $[-S,S]$.
\end{itemize}

Our scheme is based on the quantum version of the characteristic function, the Moyal quasicharacteristic function,  
or quantum characteristic function. 
Recall that for a classical probability distribution $\P(\xi)$, one can define its characteristic function as the Fourier transform 
\begin{equation} 
\Z(\chi)=\int d\xi e^{i\chi \xi} \P(\xi) .
\end{equation}
The derivatives of $\Z$ at $\chi=0$ give the moments of the distribution, its logarithmic derivatives give the cumulants 
\cite{Lukacs1960}. 
For a classical point-like particle in one dimension, $\xi=(p,q)$, momentum and position. 
In quantum mechanics, however, the momentum and position operators $\Hat{p}$ and $\Hat{q}$, 
do not commute, hence it is not possible, in general, 
to characterize a quantum point-like particle in one dimension through a nonnegative probability $\P$. 
Instead, we must use the Wigner function $\W(p,q)$, which can take negative values. 
The quantum characteristic function $\M$ is then defined as the Fourier transform of the Wigner function, 
$\M(\chi_p,\chi_q)=\int dpdq \exp{[i\chi_p p+i\chi_q q]} \W(p,q)$. After 
some straightforward algebra, 
\begin{align}
&\M(\chi_p,\chi_q)=\langle \exp[i\chi_p \Hat{p}+i\chi_q \Hat{q}]\rangle
,
\label{eq:defm}
\end{align} 
where the quantum mechanical average is defined as 
\begin{equation}
\langle\Hat{O}\rangle
=\Tr[\Hat{O}\rho],
\label{eq:qav}
\end{equation}
with $\rho$ the density operator and $\Tr$ the trace. 
The quantum characteristic function is obtained thus by the inverse Weyl-Wigner transform \cite{Weyl1928,Wigner1932}. 
It solves the question: given the classical moments $\overline{p^m q^n}$, what is the equivalent 
quantum mechanical expression in terms of averages \eqref{eq:qav}? 
In the simple case $\overline{pq}$ we know that the prescription is to take the symmetric combination 
$\langle\Hat{p}\Hat{q}+\Hat{q}\Hat{p}\rangle/2$, but for higher powers there are several possible 
combinations. 
As it turns out, the correct combination of $\langle\Hat{p}\dots\Hat{q}\dots\rangle$ is obtained by differentiating 
the Moyal function at $\chi_p=0,\chi_q=0$. This is equivalent to taking the average with the Wigner function 
$\overline{p^mq^n} \to \int dpdq p^mq^n \W(p,q)$.

Now, for a finite-dimensional system, the questions arise, 
\begin{enumerate}
\item
\emph{How do we define two complementary operators $\Hat{A}$ and $\Hat{B}$?} 
\item
\emph{How do we define the quantum characteristic function?} 
\end{enumerate}
Clearly, we put the restriction that in the limit $d\to\infty$ of an infinite dimension, $\Hat{A}\to \Hat{q}$, $\Hat{B}\to \Hat{p}$, and the definition \eqref{eq:defm} is recovered. 
The answers to the questions above are not unique, since the quantum characteristic function \eqref{eq:defm} can 
be written in several equivalent ways using the Baker-Campbell-Hausdorff formula, making the 
extension to a finite dimension ambiguous. 
The sense in which the operators $\Hat{A}$ and $\Hat{B}$ are complementary 
cannot be that a relation $[\Hat{A},\Hat{B}]=i$ is satisfied, 
since, by taking the trace of this expression we get the contradiction $0=i\,d$. 
The canonical commutation relation can be obeyed only in an infinite-dimensional space, where the domain of 
$\Hat{q}$ and $\Hat{p}$ is a proper subset of the full Hilbert space. 
Question 2 is strictly related to the generalization of the Wigner function to a finite-dimensional system, 
a subject of great interest that has spawned many different proposals \cite{Ferrie2011}. 

Here, instead of extolling the virtues of our pet proposal based on 
aesthetic considerations, we take a pragmatic attitude: we consider the sequential measurement of two arbitrary operators, then 
define the pair of observables $\Hat{A},\Hat{B}$ as complementary when they simplify the expression for the measurement, and define the discrete characteristic function in 
such a way that the final characteristic function of the outputs has a simple expression in terms of it as well. 
The definitions presented below, hence, were not chosen arbitrarily, but were suggested by the physics, as 
explained in the Methods section. 

We answer question 1 following Schwinger \cite{Schwinger1960}: 
we consider an orthonormal basis $|m\rangle$ labelled by an index $m\in \mathcal{I}=\{-S,-S+1,\dots,S\}$, 
with $d=2S+1$ the dimension of the Hilbert space. Thus $m$ is either an integer (a half-even) or a half-odd number, depending which of the two $S$ is.  
We define the conjugate basis as 
\begin{equation}
|\widetilde{m}\rangle =\frac{1}{\sqrt{d}}\sum_{m'} {\exp[2\pi i m m'/d]} |m'\rangle .
\end{equation}
It is easy to check that $|\widetilde{m}\rangle$ form an orthonormal basis when $m$ ranges in $\mathcal{I}$.
Notice that the tilde symbol is associated to the basis, not to the index $m$. 

\begin{figure}
\includegraphics[width=6in]{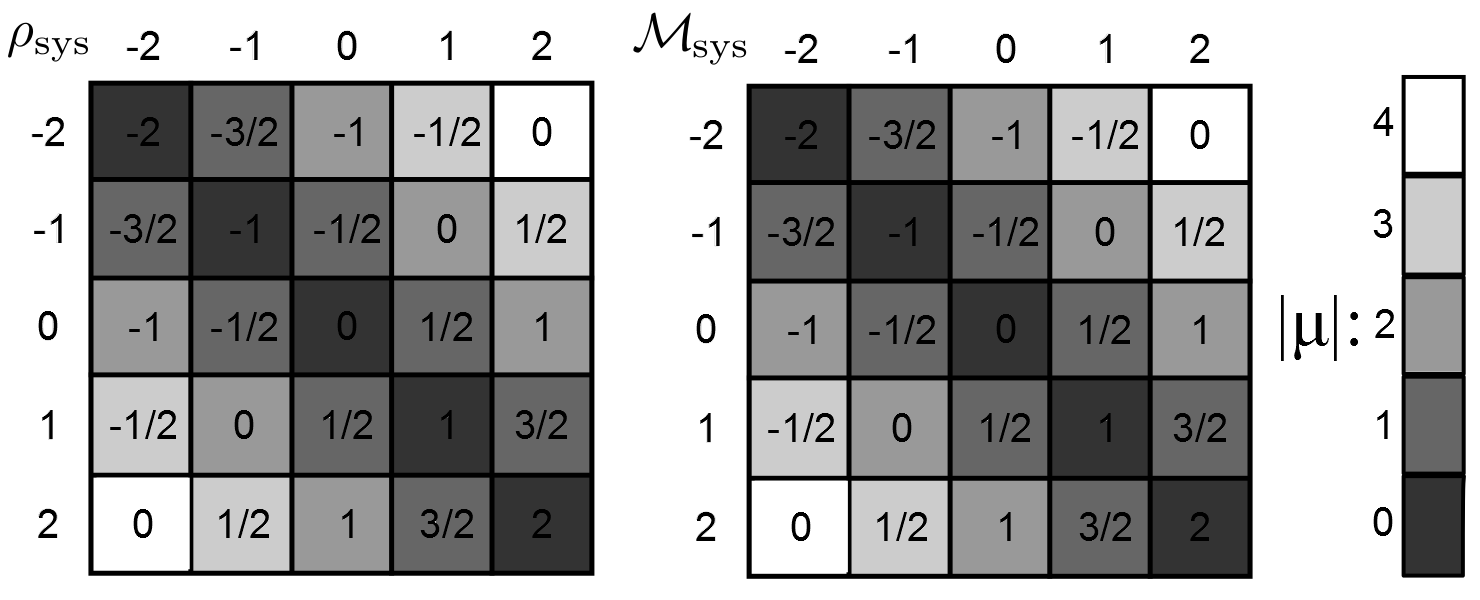}
\caption{\label{fig:ex} Allowed values of $\bar{M}$ (numbers) for fixed $\mu$ (color) in $d=5$. Elements of $\M_\mathrm{sys}$ 
with a given $\mu$ are combinations of the elements of $\rho_\mathrm{sys}$ with the same $\mu$, i.e. belonging 
to the same diagonal parallel to the main diagonal of the density matrix.}
\end{figure}

We define an operator $\Hat{A}$ having $m a_0$ as eigenvalues and $|m\rangle$ as eigenstates, and 
an operator $\Hat{B}$ having the eigenvalues $m b_0$ but $|\widetilde{m}\rangle$ as eigenstates; 
the scales are $a_0=l_0/\sqrt{d}$ and $b_0=2\pi/(l_0\sqrt{d})=2\pi/(d a_0)$, with $l_0$ some fundamental length scale. 
The scaling factors guarantee that $\Hat{A}\to\Hat{q}$ and $\Hat{B}\to\Hat{p}$ for $d\to\infty$. 
We consider a sequential measurement, with a first probe measuring $\Hat{A}$, and then 
a second probe measuring $\Hat{B}$.
Here and in the following, we consider the momentum $p$ in units of $\hbar$, so that it has dimensions $\mathrm{L}^{-1}$. 
We remark that 
\begin{equation}
\exp[i z a_0\Hat{B}]|m\rangle =(-1)^{(d-1) r_{m-z}} |f(m-z)\rangle
\end{equation}
 for any $m\in\mathcal{I}$ and any $z\in\mathbb{Z}$, 
where $f(m-z)$ is the difference $m-z$ reduced to the interval $\mathcal{I}$ by subtracting an appropriate integer multiple of $d$, $r_{m-z} d$.  
In particular, $\exp{[i a_0 \Hat{B}]}|\!-\!S\rangle=(-1)^{d-1} |S\rangle$ 
and $\exp{[-i a_0 \Hat{B}]}|S\rangle=(-1)^{d-1} |\!\!-\!\!S\rangle$. 
Thus $\Hat{B}$ is the generator of the modular translations for 
the basis $|A\rangle$.  The viceversa also holds true. 
As a matter of fact, our conventions differ from the ones used by Schwinger \cite{Schwinger1960}, and coincide 
with the ones introduced by de la Torre and Goyeneche \cite{Torre2003}.

We answer question 2 defining the Moyal function as 
\begin{equation}
\M_\mathrm{sys}(\phi_A;a)=\sum_{\bar{M}} e^{i\phi_A \bar{M} a_0}
\langle \bar{M}+\frac{\mu}{2}|\rho_\mathrm{sys}
|\bar{M}-\frac{\mu}{2}\rangle ,
\label{eq:defmsys}
\end{equation} 
where $\mu$ is an integer of the form $m-m'$, with $m,m'\in \mathcal{I}$, so $\mu\in[1-d,d-1]$, and $a=\mu a_0$. 
The sum over $\bar{M}$ is restricted by the condition that $\bar{M}\pm \mu/2$ belong to $\mathcal{I}$. 
See Fig.~\ref{fig:ex} for an example. 
This is a fundamental difference from the definition proposed by Leonhardt \cite{Leonhardt1995b,Leonhardt1996}. 
For instance, if $\mu$ takes its maximum value $\mu=2S=d-1$, then $\bar{M}$ can only be zero. 
In general, the values of $\bar{M}$ go from $-S+|\mu|/2$ to $S-|\mu|/2$, and $\bar{M}$ is integer or half-odd 
depending whether $S-|\mu|/2$ is. 
While the Moyal function Eq.~\eqref{eq:defmsys} is defined for any $\phi_A$, in order to invert it 
we need to evaluate only at the finite discrete values $\phi_A= 2\pi \bar{M}_A /[a_0(d-|\mu|)]$, 
with $\bar{M}_A\in[-S+|\mu|/2,S-|\mu|/2]$, 
\begin{equation}
\langle m|\rho_\mathrm{sys}|m'\rangle  
=\sum_{\bar{M}_A}
\frac{e^{-2\pi i\bar{M}_A \bar{M}/(d-|\mu|)}}{d-|\mu|} 
\M_\mathrm{sys}\left(\frac{2\pi \bar{M}_A}{a_0(d-|\mu|)};\mu a_0\right),
\label{eq:invrel}
\end{equation}
with $\bar{M}=(m+m')/2$ and $\mu=m-m'$.

As an example, consider a spin-1/2 particle. Then we can take $\Hat{A}=\sigma_z/2$, 
and $\Hat{B}=-\pi\sigma_y/2$ as complementary observables, with $\sigma_j$ Pauli matrices, 
having chosen $l_0=\sqrt{2}$ and hence $a_0=1$, $b_0=\pi$. 
The general state $\rho_\mathrm{sys}=(1+\mathbf{n}\cdot\boldsymbol{\sigma})/2$ has the characteristic function 
$\M_\mathrm{sys}(\phi_A;0)=\cos{(\phi_A/2)}+i\sin{(\phi_A/2)} n_z$, 
$\M_\mathrm{sys}(\phi_A;\pm 1)=(n_x\mp i n_y)/2$. 
In this case, the inversion formula \eqref{eq:invrel} gives directly the off-diagonal elements for $\mu=1$, 
while for $\mu=0$ the required values of $\phi_A$ are $\pm \pi/2$.

Finally, we assume that the initial quantum state of the probes is known, that the pointer variables $\Hat{J}_A,\Hat{J}_B$, 
 have a continuous spectrum and thus have conjugate variables, $\Hat{\Phi}_A,\Hat{\Phi}_B$, respectively. 
Starting from the initial density operator of the two probes $\rho_\mathrm{pr}$, we infer their  
initial Moyal function   
\begin{align}
&\M_\mathrm{pr}(\phi;j)=\langle\exp(i\phi \cdot\Hat{J}+i j \cdot\Hat{\Phi}) \rangle=
\int\! dJ\, 
 e^{i\phi\cdot J}
\langle J+\frac{j}{2}|\,\rho_\mathrm{pr}|J-\frac{j}{2}\rangle .
\label{eq:defmpr}
\end{align} 
For brevity, we are indicating with $J=(J_A,J_B)$, $\phi=(\phi_A,\phi_B)$,etc. vectors in an auxiliary 
two-dimensional Euclidean space.

%
\section{Results}
\begin{figure}
\includegraphics[width=6in]{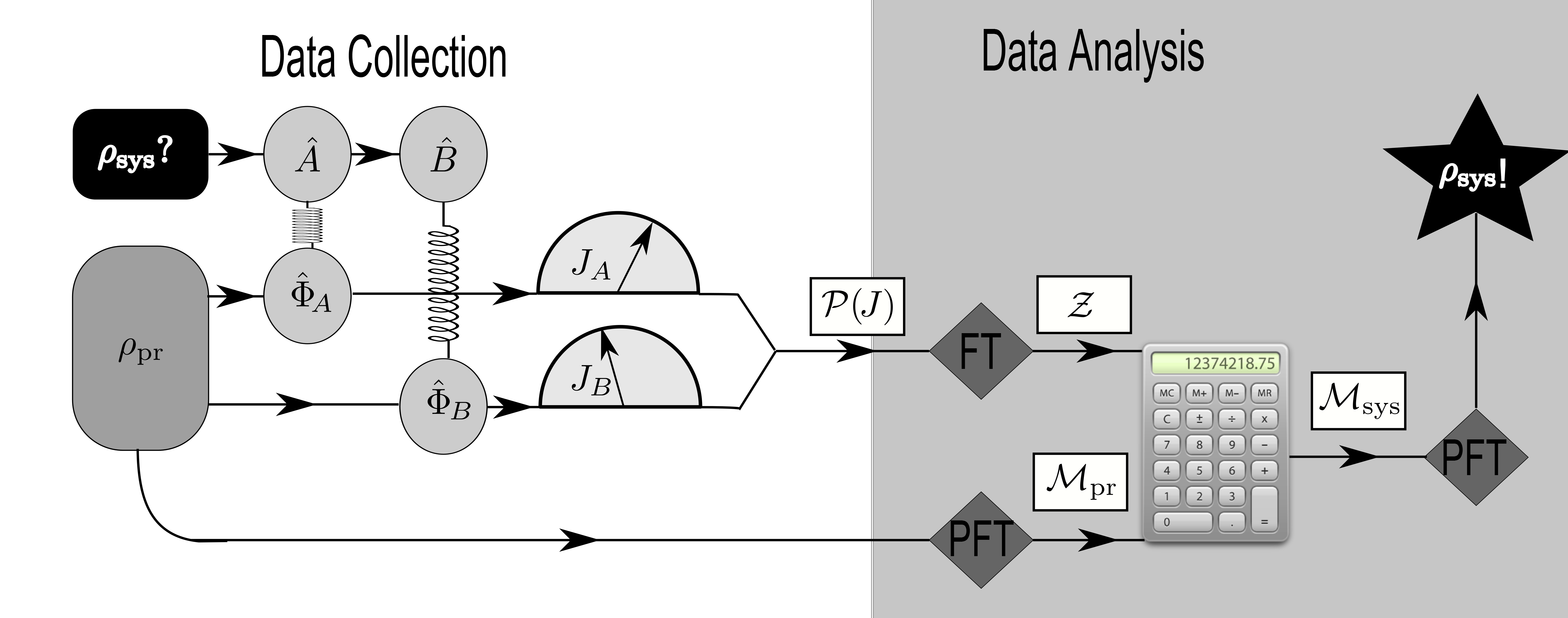}
\caption{\label{fig:scheme} Schematic diagram of the tomography. 
A system in an unknown state couples sequentially to two detectors 
through the interactions $\Hat{A} \Hat{\Phi}_A$ and $\Hat{B} \Hat{\Phi}_B$. The outputs $J_A,J_B$ 
of the probes are read, the measurement 
repeated a large number of times, and the joint probability $\mathcal{P}(J)$ estimated. 
Then, the characteristic function $\Z$ is extracted through a Fourier transform (FT), 
and the known state of the probes undergoes a partial Fourier transform (PFT) to give their Moyal 
function $\mathcal{M}_\mathrm{pr}$. Simple algebraic operations (calculator icon) 
are then applied to $\Z$ and $\M_\mathrm{pr}$ in order to get the Moyal function of the system $\M_\mathrm{sys}$. 
Finally, a partial Fourier transform yields the target density matrix.}
\end{figure}
After repeating many times the measurement of $\Hat{A}$ and $\Hat{B}$, 
we can estimate $\P(J_A,J_B)$, the joint probability of observing the outputs $J_A,J_B$ in two probes 
that make a nondemolition measurement of the system. 
Then we calculate $\Z(\phi_A,\phi_B)$, the final characteristic function, i.e., the Fourier transform of $\P(J_A,J_B)$. 
The following relation holds between the final characteristic function and the initial Moyal functions 
\begin{align}
\Z(\phi) =&\M_\mathrm{pr}(\phi;-\phi\sigma_+) \M_\mathrm{sys}(\phi_A;\phi_B)
+\M_\mathrm{pr}(\phi;-\bar{\phi}\sigma_+) \M_\mathrm{sys}(\phi_A;\bar{\phi}_B),
\label{eq:main}
\end{align} 
for any $\phi_A$ and for $\phi_B=\mu a_0$, with $\mu$ an integer in the range $[1-d,d-1]$, excluding $\mu=0$; 
here, $\sigma_+=\begin{pmatrix}0&1\\0&0\end{pmatrix}$, $\bar{\phi}=(\phi_A,\bar{\phi}_B)$, 
and $\bar{\phi}_B=[\mu-\mathrm{sgn}(\mu)d]a_0$. 
Equation~\eqref{eq:main} is the central result of this paper. 

Notice that $\bar{\bar{\phi}}_B=\phi_B$. 
Thus, if we take Eq.~\eqref{eq:main} at $\phi=(\phi_A,\bar{\phi}_B)$, we have a closed system 
of two linear equations in the two unknowns $x=\M_\mathrm{sys}(\phi_A;\phi_B)$ and 
$y=\M_\mathrm{sys}(\phi_A;\bar{\phi}_B)$.  
Therefore, we have to solve several decoupled linear equations in two unknowns for different values of 
$\phi_A$.  
This allows to finally reconstruct the density matrix in the basis 
of the eigenstates of $\Hat{A}$ by using Eq.~\eqref{eq:invrel}. Figure \ref{fig:scheme} illustrates the above procedure. 
In the limit $d\to\infty$, the second addend in Eq.~\eqref{eq:main} goes to zero, and the result of Ref.~\cite{DiLorenzo2013a} is then recovered. 
%

%

%

\section{Discussion}
An issue to consider is whether assuming the state of the detectors to be known introduce 
some circularity in the argument. On one hand, we could consider self-consistent calibration and bootstrapping, 
and on the other hand, the state of the detectors could be determined by means of a standard quantum 
state tomography scheme for a continuous variable \cite{Lvovsky2009}. Then, one would know that 
the detectors prepared in such and such way are in a state $\rho_\mathrm{pr}$, and could use them 
to apply the tomographic scheme presented herein to determine the state of any quantum system that 
couples appropriately to the detectors, leading to an overall increased efficiency.

For simplicity of exposition, we used the von Neumann model of measurement and assumed that the readout 
of the detectors had infinite precision. However, the results are valid for any non-demolition sequential measurement, 
and it can be shown that, under some hypotheses, a finite resolution in the readout introduces a factor $z_0(\phi)$ 
in front of the right hand side of Eq.~\eqref{eq:main}

\section{Methods}
Let us consider the probability of observing a readout $J=(J_A,J_B)$ from the two detectors after they have interacted with the system through the von Neumann model 
\begin{equation}
H_{int}=-\delta(t+\tau) \Hat{A}\Hat{\Phi}_A -\delta(t-\tau) \Hat{B}\Hat{\Phi}_B ,
\end{equation} 
with $\tau\to 0^+$ an infinitesimal time. 
For now, no relation is assumed between the observables of the system $\Hat{A}$ and $\Hat{B}$. 
The variables $\Hat{\Phi}$ belong to the detectors, and they are conjugated to the readout 
variables, $[\Hat{\Phi},\Hat{J}]=i$. 
By Born's rule, 
\begin{equation}
\P(J) = \Tr\left\{\left[\id\otimes \Hat{\Pi}(J)\right] U_\mathrm{int} \left[\rho_\mathrm{sys}\otimes\rho_\mathrm{pr}\right]U_\mathrm{int}^\dagger\right\}
\label{eq:genprob}
\end{equation}
with $U_\mathrm{int}=\exp[i\Hat{B}\Hat{\Phi}_B] \exp[i\Hat{A}\Hat{\Phi}_A]$ the time-evolution operator and 
$\Hat{\Pi}(J)$ the projection operator over the eigenstates of $\Hat{J}$ with eigenvalues $J$. 

Next, we consider the characteristic function, defined as the Fourier transform of the observable probability, 
\begin{equation}
\Z(\phi) = \int dJ e^{i\phi\cdot J} \P(J) = 
\Tr\left\{\left[\id\otimes e^{i\phi\cdot \Hat{J}}\right] U_\mathrm{int} \left[\rho_\mathrm{sys}\otimes\rho_\mathrm{pr}\right]U_\mathrm{int}^\dagger\right\}.
\label{eq:genchar}
\end{equation} 
We write the trace as 
\begin{equation}
\Tr[\Hat{O}]=\sum_B \int dJ \langle B,J|\Hat{O}|B,J\rangle, 
\end{equation} 
 obtaining 
\begin{align}
\Z(\phi)=&\sum_{B,A,A'}
\int dJ  e^{i\phi\cdot J}   
\langle J-C| \rho_\mathrm{pr}|J-C'\rangle 
\langle B|A \rangle 
\langle A|\rho_\mathrm{sys}| A'\rangle 
\langle A'|B \rangle ,
\label{eq:genchar3}
\end{align}
where we wrote the initial state of the system in the basis of eigenstates of $\Hat{A}$, 
$\rho_{sys}=\sum_{A,A'}|A\rangle\langle A|\rho_\mathrm{sys}| A'\rangle \langle A'|$, 
exploited the fact that $\Hat{\Phi}$ generates the translations in the $|J\rangle$ basis, 
$\exp[ix\cdot\Hat{\Phi}]|J\rangle=|J+x\rangle$, 
and defined the auxiliary vectors $C=(A,B)$, $C'=(A',B)$. 
Now, let us define $\bar{A}=(A+A')/2$ and $a=A-A'$, and change the integration variables to 
$J_A-\bar{A}$ and $J_B-B$. Then, 
\begin{align}
\Z(\phi)=&
\sum_{a}
\M_\mathrm{pr}(\phi;j_a) 
N_\mathrm{sys}(a|\phi),
\label{eq:lineqs}\\
N_\mathrm{sys}(a|\phi) =&
\sum_{\bar{A}\in D_a}\!\!\!e^{i\phi_A \bar{A}}   
\langle \bar{A}-\frac{a}{2} | e^{i\phi_B\Hat{B}}|\bar{A}+\frac{a}{2} \rangle 
\langle \bar{A}+\frac{a}{2}|\rho_\mathrm{sys}| \bar{A}-\frac{a}{2}\rangle ,
\label{eq:defnsys}
\end{align}
where $j_a=(-a,0)$ and 
we introduced the Moyal quasicharacteristic function for the probes, as defined 
in Eq.~\eqref{eq:defmpr}.
Notice that the domain of summation in $\bar{A}$ depends on $a$. 
In general, equations~\eqref{eq:lineqs} and \eqref{eq:defnsys} are too complicated to invert and be useful in reconstructing 
the quantum state. For instance, if $\Hat{A}$ and $\Hat{B}$ commute, only diagonal terms contribute to 
$N_\mathrm{sys}(a|\phi)$, so that no reconstruction of the quantum state is possible, as one can only find the the diagonal elements of $\rho_{sys}$, as expected. 
Furthermore, if $\Hat{A}$ and $\Hat{B}$ have mutually unbiased eigenbases with a constant relative phase, such that 
$\langle A|B\rangle=1/\sqrt{d}$, then
$N_\mathrm{sys}(a|\phi) =g(\phi_B) \M_\mathrm{sys}(\phi_A;a)$, with $g(\phi_B)=\sum_B \exp(i\phi_B B)/d$, and 
no actual simplification occurs. 

On the other hand, it is clear from Eq.~\eqref{eq:defnsys} that if for some $\phi_B$ the operator 
$\exp{(i\phi_B\Hat{B})}$ translates the eigenstates of $\Hat{A}$ into each other, 
then only few terms (precisely, two) in $a$ survive. 
Thus, we exploit the freedom that we have in choosing the bases $|A\rangle$ and $|B\rangle$, and we assume that 
they are Fourier conjugated, i.e.,
\begin{equation}
\langle A|B\rangle = \frac{\exp[i B A]}{\sqrt{d}},
\label{eq:fourconj}
\end{equation}
with the eigenvalues of $\Hat{A}$ being of the form $A=ma_0$, and those of $\Hat{B}$ being 
$B=m b_0$, with $m$ an integer or half-odd in the range $[-S,S]$. 

We write $\exp[i\phi_B\Hat{B}]$ in Eq.~\eqref{eq:defnsys} as 
$\sum_B |B\rangle\langle B| \exp[i\phi_B B]$, then 
substitute Eq.~\eqref{eq:fourconj} in Eq.~\eqref{eq:defnsys} so rewritten, obtaining 
\begin{align}
N_\mathrm{sys}(a|\phi) =&
\sum_m \sum_{M}\frac{e^{i\phi_A \bar{A}+i(\phi_B-a) B}}{d}   
\langle \bar{A}+\frac{a}{2}|\rho_\mathrm{sys}| \bar{A}-\frac{a}{2}\rangle 
= \frac{\sin{[\pi(\phi_B-a)\sqrt{d}]}}{d\sin{[\pi(\phi_B-a)/\sqrt{d}]}} 
\M_\mathrm{sys}(\phi_A;a),
\label{eq:defnsys2}
\end{align}
with $B=m b_0$, $m\in \mathcal{I}$, $\bar{A}=\bar{M} a_0$,  
$\bar{M}\in[-S+|\mu|/2,S-|\mu|/2]$, $a=\mu a_0$, $\mu\in[1-d,d-1]$. 
We introduced 
the Moyal quasicharacteristic function of the system, relative to the $|A\rangle$ basis, defined in Eq.~\eqref{eq:defmsys}. 
Furthermore, for $\phi_B=\mu' a_0$, $\mu'\in[1-d,d-1]$, 
$N_\mathrm{sys}(a|\phi)$ in Eq.~\eqref{eq:defnsys2} simplifies to 
\begin{align}
N_\mathrm{sys}(a|\phi)&=
\delta_{a,\phi_B} \M_\mathrm{sys}(\phi_A;\phi_B) 
+
\delta_{a,\bar{\phi}_B} \M_\mathrm{sys}(\phi_A;\bar{\phi}_B) .
\label{eq:discr}
\end{align} 
For $\phi_B=0$, instead, only one term survives, 
\begin{equation}
N_\mathrm{sys}(a|(\phi_A,0))=\delta_{a,0} \M_\mathrm{sys}(\phi_A;0)
. 
\label{eq:discr0}
\end{equation}
Hence, after substituting Eq.~\eqref{eq:discr} into Eq.~\eqref{eq:lineqs} 
evaluated at the discrete points 
$\phi_B=\mu' a_0$, we get the main result Eq.~\eqref{eq:main}.
%

\acknowledgments
\subsection{Acknowledgments}
I thank Giuseppe Falci for discussions. 
This work was performed as part of the Brazilian Instituto Nacional de Ci\^{e}ncia e
Tecnologia para a Informa\c{c}\~{a}o Qu\^{a}ntica (INCT--IQ) and it was partially funded 
by the Conselho Nacional de Desenvolvimento Cient\'{\i}fico e Tecnol\'{o}gico (CNPq) 
through process no. 245952/2012-8.

\appendix
\section{Appendix: Comparison with Leonhardt's definition of Moyal function}
Leonhardt \cite{Leonhardt1995b,Leonhardt1996} proposed a tomographic scheme based on a definition of 
quantum characteristic function for finite-dimensional Hilbert spaces. 
While Leonhardt's definition differs from ours, the two definitions are related, and in the following we shall discuss them. 
We base our discussion on Ref.~\cite{Leonhardt1996}. 

First, a remark about the notation. Leonhardt uses an index $m$ that ranges from $-S$ to $S=(d-1)/2$ for odd $d$, 
and from $1-d/2$ to $d/2$ for even $d$. We shall use the letter $l$, instead of $m$, for such an index, 
while we keep $m$ to denote an integer or half-odd in the range $[-S,S]$, as in the main text. 
Furthermore, the states $|l\rangle$ coincide with our states $|m\rangle$ for odd $d$, but for even $d$ there is a 
different notation  between us and Leonhardt. Here, we shall indicate as customary with $|m\rangle$ 
the states of the tomographic basis, with the proviso that for even $d$, Leonhardt 
uses the notation $|m+1/2\rangle_L$. To keep the notation compact, we introduce the 
number $f=1$ for even $d$ and $f=0$ for odd $d$, representing the fermionic character of the Hilbert space.

Leonhardt defines the characteristic function as 
\begin{align}
\widetilde{W}(\nu,n)=&\sum_{l=-S+f/2}^{S+f/2} \exp\left[-\frac{4\pi i}{d} n(l-\nu)\right]\sideset{_L}{}{\mathop{\langle}} l|\rho|l-2\nu\rangle_L
\nonumber
\\
=&
\sum_{m=-S}^{S} \exp\left[-\frac{4\pi i}{d} n(m+f/2-\nu)\right]\langle m|\rho|m-2\nu\rangle,
\label{eq:leo}
\end{align}
with the convention that whenever $m-2\nu$ is outside the range $[-S,S]$, it is reduced back to it by 
adding or subtracting an appropriate multiple of $d$. 
Notice that $2\nu$ is limited to integer values, but $n$ can be arbitrary. Thus, for ease of comparison, we shall 
put $-4\pi n/d=\phi$, change the order of the arguments of $\widetilde{W}$, substitute $\phi$ to $n$ as first argument, 
and $\mu$ to $2\nu$ as second argument
$\widetilde{\W}(\phi;\mu)\equiv \widetilde{W}(\mu/2,-\phi d/(4\pi))$. 
Furthermore, noticing that $\widetilde{\W}(\phi;\mu+d)=\exp[-i\phi d/2]\widetilde{\W}(\phi;\mu)$, 
the values of $\mu$ can be restricted to the range $[0,d-1]$. 

In the main article, we defined the Moyal function as 
\begin{equation}
\M(\phi;\mu)=\sum_{\bar{M}=-S+|\mu|/2}^{S-|\mu|/2} e^{i\phi \bar{M}}\langle \bar{M}+\frac{\mu}{2}|\rho
|\bar{M}-\frac{\mu}{2}\rangle .
\label{eq:moy}
\end{equation} 
With the position $m\to m-\mu/2$, we can rewrite Eq.~\eqref{eq:leo} as 
\begin{align}
\widetilde{\W}(\phi;\mu)
=&
\sum_{m=-S-\mu/2}^{S-\mu/2} \exp\left[i \phi(m+f)\right]\langle m+\mu/2|\rho|m-\mu/2\rangle.
\label{eq:leo2}
\end{align}
For $\mu=0$, we have that the two definitions coincide, apart from a phase factor for the even dimensional case, 
\begin{equation}
\widetilde{\W}(\phi;0)=e^{i\phi f/2}\M(\phi;0).
\end{equation} 
For $\mu>0$, we can split the sum in Eq.~\eqref{eq:leo2} as 
\begin{align}
\widetilde{\W}(\phi;\mu)
=&
\left[\sum_{m=-S+\mu/2}^{S-\mu/2}+\sum_{-S-\mu/2}^{-S+\mu/2-1}\right] \exp\left[i \phi(m+f/2)\right]
\langle m+\mu/2|\rho|m-\mu/2\rangle.
\label{eq:leo3}
\end{align}
The first sum yields $\exp[i\phi f/2]\M(\phi;\mu)$. 
In the second sum, we put $\mu=\widetilde{\mu}+d$ (notice that $\widetilde{\mu}<0$), $m=\widetilde{m}-d/2$, obtaining 
\begin{equation}
\sum_{\widetilde{m}=-S-\widetilde{\mu}/2}^{S+\widetilde{\mu}/2} \exp\left\{i \phi[\widetilde{m}+(f-d)/2]\right\}
\langle \widetilde{m}+\widetilde{\mu}/2+d|\rho|\widetilde{m}-\widetilde{\mu}/2\rangle=
\exp[i\phi (f-d)/2]
\M(\phi;\widetilde{\mu}),
\end{equation}
where we exploited the convention that $|\widetilde{m}+\widetilde{\mu}/2+d\rangle=
|\widetilde{m}+\widetilde{\mu}/2\rangle$. 
Thus, we find that for $\mu>0$, 
\begin{equation}
\widetilde{\W}(\phi;\mu) =\exp[i\phi f/2]\M(\phi;\mu)+\exp[i\phi (f-d)/2]
\M(\phi;\mu-d). 
\end{equation}
In particular, for the discrete values considered by Leonhardt, $\phi=-4\pi n/d$, 
\begin{equation}
\widetilde{W}(\nu,n) =\exp[-2\pi i n f/d]\left[ \M(-4\pi n/d;2\nu)+
\M(-4\pi n/d;2\nu-d)\right]. 
\end{equation}

\section{Appendix: Further discussion}
As both $\Hat{A}$ and $\Hat{B}$ have the same eigenvalues, except for a trivial rescaling, we can write 
$\Hat{B}=(b_0/a_0) U\Hat{A}U^\dagger$, with $U$ a unitary operator.  
Precisely, 
$U= \sum_m |\widetilde{m}\rangle\langle m|$. 
Let us say, for definiteness, that the Hilbert space represents an angular momentum $S$, and that 
$\Hat{A}=a_0 \Hat{S}_z$ is proportional to an angular momentum operator, 
in the sense that upon rotation it transforms accordingly. 
The natural question arises: is $\Hat{B}/b_0$ an angular momentum operator as well? i.e., is there a 
unit vector $\mathbf{n}$ such that $\Hat{B}=b_0 \mathbf{n}\cdot\Hat{\mathbf{S}}$?
The answer is no, unless $d=2$, since in this latter case any unitary operator corresponds to a rotation. 
In general, however, the distinct unitary operators, modulo a global phase, are characterized by $d^2-1$ real parameters, 
while there are only three independent rotations \footnote{An important side question is then: 
what physical operations do the remaining $d^2-4$ generators represent? If the angular momentum is obtained adding several spin 1/2 systems, 
we know that these additional operations are individual spin rotations and entangling transformations. 
This consideration implies that all elementary particles are either spin 1/2 fermions, spin 0 bosons, or spin 1 massless bosons.}. 
The proof that, for $d>2$, none of these rotations yields $\Hat{B}/b_0=\Hat{S}_\mathbf{n}$ is as follows: 
since $\exp(-(i/\sqrt{d})\Hat{B})|S\rangle=\pm |-S\rangle$, $\Hat{B}$ must be 
$\Hat{B}=(2z+1)\pi\sqrt{d} \Hat{S}_\perp$, with 
$z\in\mathbb{Z}$ and the 
$\perp$ symbol indicating an appropriate direction in the plane orthogonal to $Z$. 
Thus, $\Hat{S}_\mathbf{n} = [(2z+1)d/2] \Hat{S}_\perp$. This equation 
implies necessarily that $\perp=\pm\mathbf{n}$, $d=2$ and either $z=0$ or $z=-1$.

Anyhow, Reck \emph{et al.} \cite{Reck1994} have proved that any unitary operator $U$ in a finite-dimensional 
Hilbert space can be realized by a suitable combination of elementary unitary operators that act nontrivially 
only in a two--dimensional subspace. Furthermore, in quantum computation, it is well known that if 
the Hilbert space is made up of $N$ distinguishable qubits, any unitary operator can be approximated at will 
by a sequence of controlled nots and of elementary unitary operations on each qubit. 

The main problem consists then in constructing the operator $\Hat{A}$, in the worst case scenario 
that this is not provided to us by Nature. 
For a system composed of $n$ distinguishable qubits, the operator $\Hat{A}$ 
can be constructed, apart from a trivial shift and rescaling as $\Hat{A}=\sum_{p=1}^N 2^{p-1}\sigma_{z,p}$, 
with $\sigma_{z,p}$ a spin operator on the $p$-th qubit. 

%
\bibliography{../weakmeasbiblio}

\end{document}